\documentclass[12pt]{article}
\usepackage{amssymb,amsmath,epsfig}
 \allowdisplaybreaks

\begin{document}
\title{\bf Effects of Electromagnetic Field on the
Structure of Massive Compact Objects}
\author{M. Z. Bhatti \thanks{mzaeem.math@pu.edu.pk} and Z. Tariq
\thanks{zohatariq24@yahoo.com}\\
Department of Mathematics, University of the Punjab,\\
Quaid-e-Azam Campus, Lahore-54590, Pakistan.}
\date{}
\maketitle
\begin{abstract}
This paper encompasses a set of stellar equations that administer
the formation and evolution of self-gravitating, dissipative
spherically symmetric fluid distributions having anisotropic
stresses in the presence of electromagnetic field. The Riemann tensor is
split orthogonally to procure five scalar functions named as
structure scalars which are then utilized in the stellar
equations. It is shown that some basic fluid properties such as
energy density inhomogeneity, pressure anisotropy and heat flux are
interlinked with the obtained scalars. Further, it is shown that all
the solutions to Einstein equations can be written in terms of these
five scalars keeping in view the static case.
\end{abstract}

\section{Introduction}

The constituents of enormous stellar configurations are tied together by
a combined force of gravitation which act on the system as a whole.
Such a gravitational force is known as self-gravitational force and
the celestial system is known as a self-gravitating system.
Self-gravitational forces play pivotal role in the comprehension of
star formation as they provide help to understand the time
evolution of stellar systems. All the celestial bodies like stars,
galaxies and galaxy clusters would disintegrate and expand without
this force. Self-gravitating fluids are classified with the help of
certain physical variables like density, pressure etc. Such fluids
have diversity of applications in relativistic astrophysics and cosmology.

Fundamental scalar functions that helps in delineating the
formation and evolution of self-gravitating fluid distributions are
termed as structure scalars. They appear when the Riemann tensor is
split orthogonally and are found to be interrelated with basic fluid
attributes including energy density inhomogeneity, pressure
anisotropy and dissipative flux etc. Such scalar functions have
irrefutable physical significance and are proved to be one of the
best ways, so far, to delineate the evolution of self-gravitating
stellar configurations.

One of the main targets in modern astrophysics  is to comprehend
the consequences of electromagnetic field in the
formation and evolution of the stellar configurations. Arba\~{n}il and Zanchin
\cite{1} analyzed the equilibrium configurations of static charged
and uncharged spheres comprising relativistic polytropic fluid. They
compared the obtained results with the spheres that have
non-relativistic polytropic fluid. Weber \cite{2} over-viewed the
astrophysical phenomena related to strange quark matter. He
discussed the possible observations associated with the states of
matter inside compact stars. Negreiros et al. \cite{3} showed that
the electric field strength may increase as strange matter formulates
a color super-conductor. Ivanov \cite{4} studied
Reissner-Nordstr\"{o}m solution by considering the perfect
fluid and found general formulas for de-Sitter solutions in the
presence of electromagnetic field.

To gain insight into the time evolution of different celestial
objects, structure scalars come in handy. A lot of work has been done by
relativists in this regard. Herrera et al. \cite{5}
considered the extension of Lemaitre-Tolman-Bondi (LTB) space-times
for the dissipative case and presented one of the symmetric
properties of LTB. They also described LTB model by using certain
scalar functions known as structure scalars. Herrera et al. \cite{6}
re-obtained the stellar evolution equations for dissipative,
anisotropic spherically symmetric fluid and analyzed the condition
for stability of shear-free condition. Herrera et al. \cite{7}
explored the physical meanings of scalar functions for dust cloud by
including the cosmological constant and investigated the changes produced
by certain factors in the inhomogeneity factor. Herrera et al.
\cite{8} worked out a set of equations for spherically symmetric
self-gravitating dissipative fluids in terms of five fundamental
scalar functions and presented that these are directly related to
the dissipative flux, local pressure anisotropy and energy density
etc. Herrera et al. \cite{9} deployed a set of equations that govern
dissipative spherically symmetric fluids. They emphasized on the
relationship between the Weyl and shear tensors, the anisotropic factor
and the density inhomogeneity.

Much efforts have been put forward by researchers in order to grasp
the concept of complexity in the stellar systems. Calbet and
L\'{o}pez-Ruiz \cite{10} derived the evolution equations for
tetrahedral gas and presented that complexity does not exceed its
minimum and maximum values. Lopez-Ruiz et al. \cite{11} proposed a
measure of complexity using probabilistic approach and showed that
it is applicable to a variety of physical situations. Herrera et al.
\cite{12} extended the notion of complexity of relativistic fluid
distributions to the vacuum solution of the Bondi metric and found a
link between vorticity and complexity. Herrera \cite{13} propounded
a novel notion for complexity of static spherically symmetric
self-gravitating systems in the framework of general relativity.
They also formulated the Einstein's equations that fulfill the
criteria of zero complexity. Crutchfield and Young \cite{14} used statistical
mechanics to delineate the complexity of non-linear mechanical
systems.

The incorporation of heat dissipative flux is mandatory for the
understanding of internal constitution and high temperature of
stars. Researchers presented different solutions for different types
of stellar systems including the effects of heat flux. Grammenos
\cite{15} presented a thermodynamical solution of Friedmann-like
spherical stellar configuration for non-adiabatic collapse.
Herrera et al. \cite{16} analyzed the outcomes of thermal conduction
within a relativistic fluid and concluded that its evolution depends
directly on thermodynamical variables. Blandford et al.
\cite{17} showed that thermal effects in the outer crust of a
neutron star give rise to magnetic field. de Oliveira et al.
\cite{18} propounded a collapsing radiating star model comprising
isotropic shear-free fluid having heat flow in radial direction.

The exploration of the solutions of Einstein equations with different background metrics and matter fields
have attracted the attention of many researchers \cite{zz}. Turimov et al. \cite{19}
proposed static axisymmetric solutions of Einstein field equations
coupled with the static and axisymmetric phantom field. They found
the validity of null energy condition for the case of phantom
field while for the case of fundamental scalar field, it does not hold. Quevedo
\cite{20} inspected the Newtonian and relativistic multipole moments
and concluded that the gravitational field of static and
axisymmetric mass distributions can be delineated using this
approach. Tolman \cite{21} developed a scheme to obtain explicit
solutions of the field equations.

The expansion scalar and shear stress tensors are used for better
understanding of the behavior of fluids. A lot of research has been
carried out using these tensors in combination with other fluid
properties. Naidu et al. \cite{22} analyzed the consequences of
anisotropic pressure and heat flux on radiating spherically
symmetric stars during its gravitational collapse. They concluded
that the relaxation time for heat flux and shear stresses are
significantly different. Ivanov \cite{23} established a general form
for the collapse of a charged spherical body having anisotropic
stresses with shear and bulk viscosities. Herrera et al. \cite{24}
proposed expansion-free spherically symmetric fluid distributions.
They presented the set of field equations as well as junction
conditions for the case of anisotropic dissipative fluid
configurations and concluded that cavity must be produced if the
evolution is expansion-free. Herrera et al. \cite{25} investigated
the attributes of dissipative axially symmetric stellar structures
with shear-free condition and concluded that the system progresses
towards FLRW space-time in the absence of dissipation. The stability
of relativistic interiors as well as the existence of
self-gravitating structures has been discussed widely in literature
\cite{25a}.

The key objective of this work is to investigate the part played by
the structure scalars in the formation and evolution of charged
self-gravitating, dissipative spherically symmetric configurations having
anisotropic stresses. The format of the paper is as follows. Section
\textbf{2} states the Einstein equations in the presence of
electromagnetic field along with the introduction of some physical
variables which are used to delineate self-gravitating dissipative
anisotropic fluid. Section \textbf{3} focuses on the orthogonal
splitting of the Riemann tensor from which five structure scalars
are procured. Section \textbf{4} lists down several alternatives to
delineate the physical relevance of such scalar functions. In section
\textbf{5}, we abridges the whole discussion.

\section{The General Formalism}

This section enlists certain stellar equations using few physical
variables that explain the behavior of self-gravitating anisotropic
fluid under the influence of electromagnetic field \cite{8}.

\subsection{Einstein-Maxwell Equations and Kinematical Quantities}

We assume a spherically symmetric collapsing fluid having
anisotropic stresses experiencing heat dissipation. The line element
for such distribution is given by
\begin{equation}\label{1a}
ds^2= e^\nu dt^2-e^\lambda dr^2-r^2 d\theta^2-r^2 \sin^2 \theta
d\phi^2,
\end{equation}
where $\nu$ and $\lambda$ are functions of temporal and radial
coordinates. The electromagnetic energy-momentum tensor is as follows
\begin{equation}\nonumber
S_{\alpha\beta}= \frac{1}{4\pi} \left( -F^\delta_\alpha
F_{\beta\delta}+\frac{1}{4}
F^{\delta\omega}F_{\delta\omega}g_{\alpha\beta}\right).
\end{equation}
The electromagnetic field tensor symbolized by $F_{\alpha\beta}$ is
given as $F_{\alpha\beta}=\varphi_{\beta,\alpha}-\varphi_{\alpha,\beta}$ with
the four-potential denoted as $\varphi^\alpha=\varphi(r)
\delta^\alpha_0$ and the four-current density as $J^\alpha=\sigma(r)
u^\alpha$. Here, $\sigma$ depicts the charge density whereas
$\varphi$ is symbolized for scalar potential. The Einstein-Maxwell
equations are given by
\begin{equation}\nonumber
F^{\alpha\beta}_{~~;\beta}=\mu_0 J^\alpha, \quad \quad \quad
F_{[{\alpha\beta;\delta}]}=0,
\end{equation}
where magnetic permeability is denoted by $\mu_0$. The
non-zero components of the Maxwell field equations provide
the following couple of equations
\begin{equation}\label{4a}
\frac{\partial^2\varphi}{\partial r^2}+
\frac{\partial\varphi}{\partial r}\left[-
\frac{{\lambda}'}{2}-\frac{{\nu}'}{2}+
\frac{2}{r}\right]=\frac{\mu_0 \sigma
e^{\lambda+\frac{\nu}{2}}}{\sqrt{1-\omega^2}},\quad
\dot{\phi}'-\phi'\left[\frac{\dot{\nu}}{2}+\frac{\dot{\lambda}}{2}\right]=\frac{\mu_0\sigma
e^{\nu+\frac{\lambda}{2}}}{\sqrt{1-\omega^2}}.
\end{equation}
Here, primed quantities depict that the derivative is taken with respect
to the radial coordinate $r$. Integration of first equation in Eq.(\ref{4a}) yields
\begin{equation}\label{5a}
\frac{\partial\varphi}{\partial r}=\frac{s
e^{\frac{\nu+\lambda}{2}}}{r^2},\quad
\textmd{where} \quad
s=\int_0^r \frac{\mu_0 \sigma r^2 e^{\lambda/2}}{\sqrt{1-\omega^2}}
dr.
\end{equation}
The components of electromagnetic stress tensor that survive are
\begin{align}\nonumber
S_{00}=\frac{s^2 e^\nu}{8\pi r^4},\quad S_{11}= - \frac{s^2
e^\lambda}{8\pi r^4},\quad  S_{22}=\frac{s^2}{8\pi r^2},\quad
S_{33}=\frac{s^2 \sin^2\theta}{8\pi r^2}.
\end{align}
We procure the Einstein equations
$(G^\mu_\nu=\kappa \left(T^\mu_\nu+ S^\mu_\nu\right)$) to obtain
\begin{align}\label{8a}
\kappa\left(T^0_0+\frac{s^2}{8\pi
r^4}\right)&=\frac{1}{r^2}-\frac{e^{-\lambda}}{r^2}+\frac{{\lambda}'e^{-\lambda}}{r},\\\label{9a}
\kappa\left(T^1_1+\frac{s^2}{8\pi
r^4}\right)&=\frac{1}{r^2}-\frac{e^{-\lambda}}{r^2}-\frac{{\nu}'e^{-\lambda}}{r},\\\label{10a}
\kappa\left(T^2_2-\frac{s^2}{8\pi
r^4}\right)&=e^{-\nu}\left(\frac{\ddot{\lambda}}{2}
+\frac{\dot{\lambda}^2}{4}-\frac{\dot{\lambda}
\dot{\nu}}{4}\right)+e^{-\lambda}\left(\frac{{\lambda}'}{2r}-\frac{{\nu}''}{2}
-\frac{{\nu}'^2}{4}-\frac{{\nu}'}{2r}+\frac{{\lambda}'
{\nu}'}{4}\right),\\\label{11a} \kappa
T_{01}&=\frac{\dot{\lambda}}{r}.
\end{align}
The pure locally Minkowski coordinates expressed as $(\tau,x,y,z)$
are given by \cite{8}
\begin{eqnarray}\nonumber
d\tau=e^{\nu/2}dt, \quad dy=rd\theta,\quad
dx=e^{\lambda/2}dr, \quad dz=r \sin\theta d\phi.
\end{eqnarray}
The Minkowski components of energy-momentum tensor are written as
\begin{eqnarray}\nonumber
\bar{T}_0^0=T_0^0,\quad\bar{T}_1^1=T_1^1,\quad
\bar{T}_2^2=T_2^2,\quad\bar{T}_3^3=T_3^3,\quad
\bar{T}_{01}=e^{-(\nu+\lambda)/2}T_{01}.
\end{eqnarray}
From the perspective of a comoving observer, the covariant
components of the energy-momentum tensor in Minkowski coordinates as
presented earlier \cite{9} becomes
$$
\begin{pmatrix}
\rho+\varepsilon & -q-\varepsilon & 0 & 0\\
-q-\varepsilon & P_r+\varepsilon & 0 & 0\\
0 & 0 & P_\bot & 0\\
0 & 0 & 0 & P_\bot \\
\end{pmatrix},
$$
so that the Lorentz transformation yields
\begin{align}\label{12a}
\bar{T}_0^0&=T_0^0=\frac{\rho+P_r
\omega^2}{1-\omega^2}+\frac{2\omega
q}{1-\omega^2}+\frac{\varepsilon(1+\omega)}{1-\omega},\\\label{13a}
\bar{T}_1^1&=T_1^1=-\left(\frac{\rho
\omega^2+P_r}{1-\omega^2}+\frac{2\omega
q}{1-\omega^2}+\frac{\varepsilon(1+\omega)}{1-\omega}\right),\\\label{14a}
\bar{T}_{01}&=e^{-(\nu+\lambda)/2}T_{01}=-\left(\frac{\omega
e^{(\nu+\lambda)/2}(\rho+P_r)}{1-\omega^2}+\frac{qe^{(\nu+\lambda)/2}(1+\omega
^2)}{1-\omega^2}+\frac{\varepsilon(1+\omega)e^{(\nu+\lambda)/2}}{1-\omega}\right),\\\label{15a}
\bar{T}_2^2&=T_2^2=-P_\bot; \quad \bar{T}_3^3=T_3^3=-P_\bot.
\end{align}
The coordinate velocity represented by $\frac{dr}{dt}$ in the coordinate
system $(t,r,\theta,\phi)$ is linked with $\omega$ as
$\omega=e^{\frac{(\lambda-\nu)}{2}}\left(\frac{dr}{dt}\right)$.
By making use of Eqs. (\ref{12a})-(\ref{15a}) into Eqs.(\ref{8a})-(\ref{11a}), we acquire
\begin{align}\nonumber
&\frac{\rho+P_r \omega^2}{1-\omega^2}+\frac{2\omega
q}{1-\omega^2}+\frac{\varepsilon(1+\omega)}{1-\omega}+\frac{s^2}{8\pi
r^4}=\frac{1}{\kappa}\left(\frac{1}{r^2}
-\frac{e^{-\lambda}}{r^2}+\frac{{\lambda}'e^{-\lambda}}{r}\right),\\\nonumber
&-\left(\frac{\rho \omega^2+P_r}{1-\omega^2}+\frac{2\omega
q}{1-\omega^2}+\frac{\varepsilon(1+\omega)}{1-\omega}\right)+\frac{s^2}{8\pi
r^4}=\frac{1}{\kappa}\left(\frac{1}{r^2}-\frac{e^{-\lambda}}{r^2}
-\frac{{\nu}'e^{-\lambda}}{r}\right),\\\nonumber
&-P_\bot -\frac{s^2}{8\pi
r^4}=\frac{1}{\kappa}\left[e^{-\lambda}\frac{-\lambda}'\left(\frac{\ddot{\lambda}}{2}
+\frac{\dot{\lambda}^2}{4}-\frac{\dot{\lambda}
\dot{\nu}}{4}\right)+e^{-\lambda}\left(\frac{{\lambda}'}{2r}
-\frac{{\nu}''}{2}-\frac{{\nu}'^2}{4}-\frac{{\nu}'}{2r}+\frac{{\lambda}'
{\nu}'}{4}\right)\right].
\end{align}
In non-comoving coordinates, the four-velocity vector for our line
element takes the form as \cite{8}
\begin{equation}\nonumber
u^\alpha=\left(\frac{e^{-\nu/2}}{\sqrt{1-\omega^2}},\frac{\omega
e^{-\lambda/2}}{\sqrt{1-\omega^2}},0,0\right).
\end{equation}
Our next target is to carry out few kinematical quantities including
four-acceleration, shear tensor and expansion scalar similar to
those found in \cite{8}. The non-zero components of
four-acceleration as given in \cite{27} in this non-comoving
coordinate system are
\begin{equation}\label{21a}
a_1\omega=-a_0e^{(\lambda-\nu)/2}=-\left(\frac{\omega^2
\omega'}{(1-\omega^2)^2}+\frac{\omega\nu'}{2(1-\omega^2)^2}+
\frac{\omega^2\dot{\lambda}e^{(\lambda-\nu)/2}}{2(1-\omega^2)}+\frac{\omega
\dot{\omega}e^{(\lambda-\nu)/2}}{(1-\omega^2)^2} \right).
\end{equation}
The shear tensor $\sigma_{\alpha\beta}$ is defined as
\begin{equation}\nonumber
\sigma_{\alpha\beta}=u_{\alpha;\beta}+u_{\beta;\alpha}-u_\alpha
a_\beta-u_\beta a_\alpha-\frac{2\Theta h_{\alpha\beta}}{3},
\end{equation}
where
\begin{equation}\nonumber
h_{\alpha\beta}=g_{\alpha\beta}-u_\alpha u_\beta,\quad \textmd{and} \quad
\Theta=u^\alpha_{~;\alpha}.
\end{equation}
The expansion scalar $(\Theta)$ takes the following value \cite{8}
\begin{equation}\nonumber
\Theta=\frac{\dot{\lambda}e^{-\nu/2}}{2\sqrt{(1-\omega^2)}}+
\frac{\omega\dot{\omega}e^{-\nu/2}}{(1-\omega^2)^{3/2}}+
\frac{\omega\nu'e^{-\lambda/2}}{2\sqrt{(1-\omega^2)}}+
\frac{\omega'e^{-\lambda/2}}{(1-\omega^2)^{3/2}}+ \frac{2\omega
e^{-\lambda/2}}{r\sqrt{(1-\omega^2)}}.
\end{equation}
The non-vanishing components of shear tensor are procured as
\cite{8}
\begin{align}\nonumber
\sigma_{00}&=\omega^2 e^{\nu-\lambda} \sigma_{11},\quad
\sigma_{01}=-\omega e^{(\nu-\lambda)/2}\sigma_{11},\\\nonumber
\sigma_{11}&=-\frac{2}{3(1-\omega^2)^{3/2}}\left[\dot{\lambda}e^\lambda
e^{-\nu/2}+\frac{2\omega\dot{\omega} e^\lambda
e^{-\nu/2}}{(1-\omega^2)}+\omega\nu'e^{\lambda/2}+\frac{2\omega'e^{\lambda/2}}{(1-\omega^2)}-\frac{2\omega
e^{\lambda/2}}{r} \right],
\\\nonumber \sigma_{22}&=-\frac{r^2e^{-\lambda}(1-\omega^2)}{2}
\sigma_{11},\quad
\sigma_{33}=-\frac{r^2e^{-\lambda}(1-\omega^2)}{2} \sin^2\theta
\sigma_{11}.
\end{align}
The shear tensor can be written in an an alternate form as \cite{8}
\begin{equation}\nonumber
\sigma_{\alpha\beta}=\frac{\sigma}{2} \left(s_\alpha
s_\beta+\frac{1}{3}h_{\alpha\beta} \right),
\end{equation}
where
\begin{equation}\nonumber
\sigma=-\frac{1}{\sqrt{(1-\omega^2)}}\left[\dot{\lambda}e^{-\nu/2}+
\frac{2\omega\dot{\omega}e^{-\nu/2}}{(1-\omega^2)}+\omega\nu'
e^{-\lambda/2}+\frac{2\omega'
e^{-\lambda/2}}{(1-\omega^2)}-\frac{2\omega e^{-\lambda/2}}{r}
\right].
\end{equation}
and
\begin{equation}\label{22a}
s^\alpha=\left(\frac{\omega
e^{-\nu/2}}{\sqrt{(1-\omega^2)}},\frac{e^{-\lambda/2}}{\sqrt{(1-\omega^2)}},0,0\right),
\end{equation}
which fulfills the following properties
\begin{equation}\nonumber
s^\alpha u_\alpha=0, \quad s^\alpha s_\alpha=-1.
\end{equation}
The energy-momentum tensor for non-comoving coordinate system is considered to be anisotropic
and dissipative as
\begin{equation}\label{23a}
T^\alpha_\beta=\tilde{\rho}u^\alpha u_\beta -\check{P}h^\alpha_\beta
+\Pi^\alpha_\beta + \tilde{q}(s^\alpha u_\beta+s_\beta u^\alpha),
\end{equation}
with
\begin{align}\nonumber
\Pi^\alpha_\beta&=\Pi\left(s^\alpha
s_\beta+\frac{1}{3}h^\alpha_\beta \right),\quad
\check{P}=\frac{\tilde{P}_r +2P_\bot}{3},\quad\Pi=\tilde{P}_r-P_\bot,\quad
\tilde{q}^\alpha=\tilde{q}s^\alpha, \quad
\\\nonumber \tilde{q}&=q+\varepsilon,\quad \tilde{\rho}=\rho+\varepsilon, \quad
\tilde{P_r}=P_r+\varepsilon.
\end{align}
To procure the junction conditions, we assume charged-Vaidya
spacetime as the exterior metric which is given by
\begin{equation}\nonumber
ds^2=\left( \frac{1-2M(u)}{R}+\frac{Q^2}{R^2}\right) du^2 +2dudR
-R^2(d\theta^2+ sin^2 \theta d\phi^2),
\end{equation}
here, the retarded time coordinate is symbolized by $u$ while $R$
represents the null coordinate. For the smooth matching of two
metrics over the boundary surface $r=r_\Sigma$, we call for the
continuity of the first and second fundamental forms across the
surface. This is done using the continuity of the line elements
across the hypersurface, i.e.,
$(ds^2_-)_\Sigma=(ds^2)_\Sigma=(ds^2_+)_\Sigma$. Next, the extrinsic
curvature $k_{\alpha\beta}$ defined as
$k^{\pm}_{\alpha\beta}=-\eta^{\pm}_\sigma\left(\frac{\partial^2
\chi^{\sigma}\pm}{\partial\varepsilon^\alpha\partial\varepsilon^\beta}+\Gamma^\sigma_{\mu\nu}\frac{\partial\chi^{\mu}\pm\partial\chi^{\nu}\pm}
{\partial\varepsilon^\alpha\partial\varepsilon^\beta}\right)$ must
also be continuous for both the interior  and exterior line
elements, i.e., $k^+_{\alpha\beta}=k^-_{\alpha\beta}$ where
$\chi^\sigma$ and $\varepsilon^\sigma$ represent the coordinates of
$\Sigma$ in the interior or exterior manifold and $\eta_\sigma$
depicts the components of vector normal to $\Sigma$. Following this
procedure, certain conditions are obtained that are considered to be
necessary and sufficient for smooth matching of exterior and
interior regions of a particular astrophysical object. Consequently,
we procure
\begin{align}\label{25a}
e^{\nu}&\overset{\Sigma}=\left(1-\frac{2M}{R}+\frac{Q^2}{R^2}\right),
\quad
e^{-\lambda}\overset{\Sigma}=\left(1-\frac{2M}{R}+\frac{Q^2}{R^2}\right),\quad
Q\overset{\Sigma}=s,
\quad P_r\overset{\Sigma}=q,
\end{align}
where $\Sigma$ specifies that the manipulation has been made over the boundary surface.
We noticed that in the absence
of dissipative flux, the radial pressure vanishes.

\subsection{The Riemann and the Weyl Tensor}

The Riemann tensor utilizes Weyl tensor $C^\rho_{\alpha\beta\mu}$,
the Ricci tensor $R_{\alpha\beta}$ and the scalar curvature $R$
whose mathematical form is given by \cite{8}
\begin{equation}\label{27a}
R^\rho_{\alpha\beta\mu}=C^\rho_{\alpha\beta\mu}+\frac{1}{2}R^\rho_\beta
g_{\alpha\mu}+\frac{1}{2}R_{\alpha\beta}\delta^\rho_\mu+\frac{1}{2}R_{\alpha\mu}\delta^\rho_\beta-\frac{1}{2}R^\rho_\mu
g_{\alpha\beta}-\frac{1}{6}R\left(\delta^\rho_\beta
g_{\alpha\mu}-g_{\alpha\beta}\delta^\rho_\mu\right).
\end{equation}
The Magnetic part for the Weyl tensor becomes zero because of
spherical symmetry. Thus, the Weyl tensor can be completely
expressed using its electric part
$E_{\alpha\beta}=C_{\alpha\beta\gamma\delta}u^\gamma u^\delta$ as
\begin{equation}\nonumber
C_{\mu\nu\kappa\lambda}=E^{\beta\delta}u^\alpha u^\gamma
(g_{\mu\nu\alpha\beta}
g_{\kappa\lambda\gamma\delta}-\eta_{\mu\nu\alpha\beta}
\eta_{\kappa\lambda\gamma\delta}),
\end{equation}
with $g_{\mu\nu\alpha\beta}$ being equal to
$g_{\mu\nu\alpha\beta}=g_{\mu\alpha}g_{\nu\beta}-g_{\mu\beta}g_{\nu\alpha}$
and $\eta_{\mu\nu\alpha\beta}$ symbolizes the Levi-Civita tensor.
We noticed that $E_{\alpha\beta}$ can also be expressed as
\begin{equation}\label{28a}
E_{\alpha\beta}=E\left(s_\alpha s_\beta+\frac{1}{3} h_{\alpha\beta}
\right),
\end{equation}
where the value for $E$ already calculated in \cite{8} is
\begin{align}\label{29a}
E=\frac{\ddot{\lambda}e^{-\nu}}{4}+\frac{\dot{\lambda}^2
e^{-\nu}}{8}-\frac{\dot{\lambda}\dot{\nu}
e^{-\nu}}{8}-\frac{{\nu}''e^{-\lambda}}{4}-\frac{\nu'^2
e^{-\lambda}}{8}+\frac{\lambda'\nu'e^{-\lambda}}{8}+\frac{\nu'e^{-\lambda}}{4r}-
\frac{\lambda'e^{-\lambda}}{4r}-\frac{e^{-\lambda}}{2r^2}+\frac{1}{2r^2}.
\end{align}
Also, the electric part of the Weyl tensor satisfies
\begin{equation}\nonumber
E^\alpha_\alpha=0, \quad E_{\alpha\gamma}=E_{(\alpha\gamma)},
\quad E_{\alpha\gamma}u^\gamma=0.
\end{equation}

\subsection{The Mass Function and the Tolman Mass}

This section incorporates two distinct and interesting definitions
of the interior mass of the spherical body under the influence of
electromagnetic field. The two masses are then interlinked using few
relations which will be used afterwards in the physical description
of the structure scalars which have influence of electromagnetic
field over them \cite{8}.

\subsubsection{The Mass Function}

The mass function $m$ for the line element given in Eq.(\ref{1a})
is defined by
\begin{equation}\label{30a}
R^3_{~232}=1-e^{-\lambda}=\frac{2m}{r}-\frac{\kappa s^2}{8\pi r^2}.
\end{equation}
Using Eqs. (\ref{27a}), (\ref{28a}) and the field equations, we get
\begin{equation}\label{31a}
\frac{3m}{r^3}=\frac{\kappa}{2}\left(\tilde{\rho}-\tilde{P_r}+\tilde{P_\bot}\right)+\frac{3\kappa
s^2}{8\pi r^4}+E.
\end{equation}
The above equation can be re-written in an alternate form as
\begin{equation}\label{32a}
m=\frac{\kappa}{2} \int^r_0 r^2\left(T^0_0+\frac{ss'}{4\pi
r^3}\right)dr.
\end{equation}
It is worth-noticing here that electromagnetism results in an
increased value of the mass function. Utilizing the above equation
along with Einstein equations, we procure another relation for mass
function as \cite{8}
\begin{equation}\label{33a}
m=\frac{\kappa r^3 }{6}\left(T^0_0+T^1_1-T^2_2\right)+\frac{\kappa
s^2}{16\pi r}+\frac{r^3E}{3}.
\end{equation}
Differentiating with respect to $r$ and utilizing Eq.(\ref{32a}),
it follows
\begin{equation}\label{34a}
\left(\frac{r^3 E}{3}\right)'=-\frac{\kappa
r^3(T^0_0)'}{6}+\left[\frac{\kappa r^3}{6}
(T^2_2-T^1_1)\right]'+\frac{\kappa s^2}{16\pi r^2}.
\end{equation}
Integrating w.r.t. $r$, we attain
\begin{equation}\label{35a}
E=-\frac{\kappa}{2r^3}\int^r_0 r^3
(T^0_0)'dr+\frac{\kappa}{2}(T^2_2-T^1_1)+\frac{3\kappa}{16\pi
r^3}\int^r_0 \frac{s^2}{r^2}dr.
\end{equation}
By substituting Eq.(\ref{35a}) in (\ref{33a}), we get
\begin{equation}\label{36a}
m(r,t)=\frac{\kappa r^3}{6}(T^0_0)-\frac{\kappa}{6}\int^r_0 r^3
(T^0_0)'dr+\frac{\kappa s^2}{16\pi r}+\int^r_0 \frac{\kappa
s^2}{16\pi r^2}dr.
\end{equation}
We can write $T^0_0=\tilde{\rho}$ and $T^1_1=\Pi$. This can only be
done if the following criteria is fulfilled \cite{8}.
\begin{enumerate}
  \item If we consider the static regime, i.e. when $\omega$ along with
all the time derivatives becomes zero.
  \item If we consider the quasistatic regime, i.e.,
$\omega^2\approx\dot{\omega}\approx\ddot{\nu}\approx\ddot{\lambda}
\approx\dot{\nu}\dot{\lambda}\approx\dot{\lambda}^2\approx0$.
  \item Just after the system leaves its state of equilibrium, i.e.
$\omega\approx\dot{\nu}\approx\dot{\lambda}\approx0$ but
$\dot{\omega}\neq0$.
\end{enumerate}
Keeping in view these three cases, Eq.(\ref{35a}) become
\begin{equation}\nonumber
E=-\frac{\kappa}{2r^3}\int^r_0 r^3 (\tilde{\rho})'dr +\frac{\kappa
\Pi}{2}+\frac{3\kappa}{16\pi r^3}\int^r_0 \frac{s^2}{r^2}dr.
\end{equation}
We can see that this equation links the Weyl tensor with pressure
anisotropy $\Pi$ and energy density inhomogeneity $\rho'$ in
addition to a charge term appearing due to the existence of
electromagnetic field. Now, Eq.(\ref{36a}) can take the following form
\begin{equation}\nonumber
m(r,t)=\frac{\kappa r^3\tilde{\rho} }{6}-\frac{\kappa}{6}\int^r_0
r^3 (\tilde{\rho})'dr+\frac{\kappa s^2}{16\pi r}+\int^r_0
\frac{\kappa s^2}{16\pi r^2}dr.
\end{equation}
This equation defines the mass function as a combination of
homogeneous distribution of energy density, the change caused by
inhomogeneity of energy density and charge terms appearing due to
electromagnetic field. We can conclude from here that the
electromagnetic field results in the increased mass of the spherical
body under consideration.

\subsubsection{Tolman Mass}

To comprehend the energy composition of a spherical body, Tolman
proposed a new definition known as Tolman mass. The mathematical
formulation as given in \cite{28} is
\begin{align}\nonumber
m_T&=\frac{\kappa}{2}\int^{r_\Sigma}_0 r^2
e^{(\nu+\lambda)/2}(T^0_0-T^1_1-2T^2_2)dr+\frac{1}{2}\int^{r_\Sigma}_0
r^2 e^{(\nu+\lambda)/2}
\frac{\partial}{\partial t}\left[\frac{\partial
L}{\frac{\partial}{\partial t}(g^{\alpha\beta}
\sqrt{-g})}\right]g^{\alpha\beta}dr,
\end{align}
where $L$ signifies the gravitational Lagrangian density. The mass
inside sphere having radius $r$ within the boundary $\Sigma$ is
\begin{align}\nonumber
m_T&=\frac{\kappa}{2}\int^r_0 r^2
e^{(\nu+\lambda)/2}(T^0_0-T^1_1-2T^2_2)dr+\frac{1}{2}\int^r_0 r^2
e^{(\nu+\lambda)/2}\frac{\partial}{\partial
t}\left[\frac{\partial L}{\frac{\partial}{\partial
t}(g^{\alpha\beta} \sqrt{-g})}\right]g^{\alpha\beta}dr.
\end{align}
After some manipulations, we acquire
\begin{align}\label{41a}
m_T=e^{(\nu+\lambda)/2}\left[m(r,t)-\frac{\kappa r^3
T^1_1}{2}+\frac{\kappa r^3}{2}\left(\frac{s^2}{8\pi r^4} \right)
\right].
\end{align}
Substituting the value of $T^1_1$ from Eq.(\ref{9a}) and $m$ from
Eq.(\ref{30a}), we attain
\begin{equation}\label{42a}
m_T=e^{(\nu+\lambda)/2}\left[\frac{\nu'e^{-\lambda}
r^2}{2}+\frac{\kappa s^2}{8\pi r} \right].
\end{equation}
It is prominent that electromagnetism results in the increased
energy budget of the system. Under the consideration of static
field, the gravitational acceleration for test particle turns out to
be \cite{8}
\begin{equation}\nonumber
a=\frac{m_Te^{-\nu/2}}{r^2}-\frac{\kappa s^2e^{\lambda/2}}{8\pi
r^3}.
\end{equation}
Working out the derivative of Eq. (\ref{42a}) with respect to $r$
and utilizing field equations along with Eqs. (\ref{33a}) and
(\ref{41a}), we acquire an alternative expression for $m_T$ as
\begin{align}\nonumber
rm_T'-3m_T&=r^3 e^{(\lambda-\nu)/2}\left[
\frac{\ddot{\lambda}}{2}+\frac{\dot{\lambda}^2}{4}-\frac{\dot{\lambda}\dot{\nu}}{4}\right]+r^3
e^{(\lambda+\nu)/2}\left[\frac{\kappa}{2}\left(T^1_1-T^2_2\right)-E
\right]\\\nonumber &+r\left(\frac{e^{(\lambda+\nu)/2}\kappa
s^2}{8\pi r}\right)'-\frac{3\kappa}{8\pi}\int_0^r
\frac{s^2e^{(\lambda+\nu)/2}}{r^2}dr.
\end{align}
We integrate the above equation and get
\begin{align}\nonumber
m_T&=(m_T)_\Sigma
\left(\frac{r}{r_\Sigma}\right)^3-r^3\int_r^{r_\Sigma}\frac{e^{(\lambda-\nu)/2}}{2r}
\left[\ddot{\lambda}+\frac{\dot{\lambda}^2}{2}-\frac{\dot{\lambda}\dot{\nu}}{2}\right]dr
-r^3\int_r^{r_\Sigma}\frac{e^{(\lambda+\nu)/2}}{r}\\\label{45a}&\times\left[\frac{\kappa}{2}(T^1_1-T^2_2)-E\right]dr
-r^3\int_r^{r_\Sigma}\frac{1}{r^3}\left(\frac{\kappa
e^{(\lambda+\nu)/2}s^2}{8\pi
r}\right)'dr+r^3\int_r^{r_\Sigma}\frac{3\kappa s^*}{8\pi r^4}dr,
\end{align}
with
\begin{equation}\nonumber
s^*=\int_0^r \frac{s^2e^{(\lambda+\nu)/2}}{r^2}dr.
\end{equation}
Substituting the value of $E$ in the above equation, we attain
\begin{align}\nonumber
m_T&=(m_T)_\Sigma
\left(\frac{r}{r_\Sigma}\right)^3-r^3\int_r^{r_\Sigma}\frac{e^{(\lambda-\nu)/2}}{2r}
\left[\ddot{\lambda}+\frac{\dot{\lambda}^2}{2}-\frac{\dot{\lambda}\dot{\nu}}{2}\right]
-r^3\int_r^{r_\Sigma}\frac{e^{(\lambda+\nu)/2}}{r}\\\nonumber
&\times\left[\frac{\kappa}{2}(T^1_1-T^2_2)+\frac{\kappa}{2r^3}\int_0^r
r^3 (T^0_0)'dr-\frac{3\kappa}{16\pi r^3}\int_0^r
\frac{s^2}{r^2}dr\right]dr -r^3\int_r^{r_\Sigma}\frac{1}{r^3}
\\\nonumber&\times\left(\frac{\kappa e^{(\lambda+\nu)/2}s^2}{8\pi
r}\right)'dr+r^3\int_r^{r_\Sigma}\frac{3\kappa s^*}{8\pi r^4}dr.
\end{align}
If we consider the three cases, which we have defined earlier, for the above equation,
it will depict that the Tolman mass for spherical body having radius $r$
within the boundary $\Sigma$ can be described using homogeneous
energy density $\rho$, local pressure anisotropy $\Pi$, changes
induced from the non-equilibrium state of the system and terms
arising due to the effect of electromagnetic field.

\subsubsection{Structure and Evolution Equations}

To gain insight into the formation and evolution of a
self-gravitating fluid, a set of stellar equations can be used which
in our case is given as \cite{8}
\begin{align}\label{47a}
&\tilde{\rho}^*+(\tilde{\rho}+\tilde{P}_r)\Theta-\frac{2}{3}\left(\Theta+\frac{\sigma}{2}\right)\Pi+
\tilde{q}^\dagger+2\tilde{q}\left(a+\frac{s^1}{r}\right)+\left(\frac{ss'}{4\pi
r^4}-\frac{s^2}{\pi r^5}\right)=0,\\\label{48a}
&\tilde{P}_r^\dagger+(\tilde{\rho}+\tilde{P}_r)a+\frac{2s^1\Pi}{r}-\frac{\tilde{q}}{3}(\sigma-4\Theta)-\tilde{q}^*=0,\\\label{49a}
&\Theta^*+\frac{\Theta^2}{3}+\frac{\sigma^2}{6}-a^\dagger-a^2-\frac{2as^1}{r}
+\frac{\kappa}{2}(\tilde{\rho}+3\tilde{P}_r)-k\Pi+\frac{\kappa
s^2}{8\pi r^4},\\\label{50a}
&\left(\frac{\sigma}{2}+\Theta\right)^\dagger=-\frac{3\sigma
s^1}{2r}+\frac{3\kappa \tilde{q}}{2},\\\label{51a}
&a^\dagger+a^2+\frac{\sigma^*}{2}+\frac{\Theta\sigma}{3}
-\frac{as^1}{r}-\frac{\sigma^2}{12}=-E-\frac{\kappa\Pi}{2}+\frac{\kappa
s^2}{8\pi r^4},\\\label{52a} &-\frac{3\kappa
s^1\tilde{q}}{2r}=\left(\Theta+\frac{\sigma}{2}\right)\left[\frac{\kappa
\tilde{P}_r}{2}+\frac{3m}{r^3}+\frac{3\kappa s^2}{8\pi
r^4}\right]+\left(E-\frac{\kappa\Pi}{2}+\kappa\tilde{\rho}+\frac{3\kappa
s^2}{16\pi r^4}\right)^*,\\\label{53a}
&\left(E+\frac{\kappa\tilde{\rho}}{2}-\frac{\kappa\Pi}{2}+\frac{\kappa
s^2}{8\pi
r^4}\right)^\dagger=\frac{3s^1}{r}\left(\frac{\kappa\Pi}{2}-E-\frac{\kappa
s^2}{8\pi
r^4}\right)+\frac{\kappa\tilde{q}}{2}\left(\frac{\sigma}{2}+\Theta
\right),\\\label{54a}
&\frac{3m}{r^3}=\frac{\kappa\tilde{\rho}}{2}+\frac{\kappa}{2}(P_\bot-P_r)+E+\frac{3\kappa
s^2}{16\pi r^4},
\end{align}
where, $f^\dagger=f_{,\mu}s^\mu$, $f^*=f_{,\mu}u^\mu$ and
$a^\mu=as^\mu$ and $\frac{\sigma}{2}+\Theta=\frac{3\omega s^1}{r}$.
All these stellar equations provide information relating to the
formation and evolution of our proposed system. Equations
(\ref{47a}) and (\ref{48a}) are the well-known conservation
equations. Equation (\ref{49a}) represents the Raychaudhuri equation
while (\ref{50a}) is derived using the Ricci identities. Utilizing
Eq. (\ref{27a}) along with the field equations, we get (\ref{51a})
and by making use of the Weyl tensor in Bianchi identities, Eqs.
(\ref{52a})and (\ref{53a}) are produced. One of the forthcoming
sections makes use of all these equations to describe the formation
and evolution of self-gravitating systems in the form of structure
scalars. As compared to the work done by Herrera et al. \cite{8},
our results show that electromagnetism adds to the complexity of the
stellar configuration.

\section{The Orthogonal Splitting of the Riemann Tensor}

Our aim is to split the Riemann tensor orthogonally to acquire
ceratin scalar quantities. For this purpose, we use the following
tensors already introduced in \cite{8} as
\begin{eqnarray}\nonumber
&&Y_{\alpha\beta}=R_{\alpha\gamma\beta\delta}u^\gamma
u^\delta,\quad Z_{\alpha\beta}=^*R_{\alpha\gamma\beta\delta}u^\gamma
u^\delta=\frac{1}{2}\eta_{\alpha\gamma\epsilon\rho}R^{\epsilon\rho}_{~~\beta\delta}u^\gamma
u^\delta,\\\nonumber
&&X_{\alpha\beta}=^*R^*_{\alpha\gamma\beta\delta}u^\gamma
u^\delta=\frac{1}{2}\eta_{~~\alpha\gamma}^{\epsilon\rho}R^*_{\epsilon\rho\beta\delta}u^\gamma
u^\delta,
\end{eqnarray}
where
\begin{equation}\nonumber
R^*_{\alpha\beta\gamma\delta}=\frac{1}{2}\eta_{\epsilon\rho\gamma\delta}R^{\epsilon\rho}_{~~\alpha\beta}.
\end{equation}
Making use of the Einstein's equations in Eq.(\ref{27a}) can be
written as
\begin{align}\label{55a}
R^{\alpha\gamma}_{~~\beta\delta}=C^{\alpha\gamma}_{~~\beta\delta}+2\kappa
T^{[\alpha}_{[\beta}\delta^{\gamma]}_{\delta]}+2\kappa
E^{[\alpha}_{[\beta}\delta^{\gamma]}_{\delta]}+\kappa
T\left[\frac{1}{3}\delta^\alpha_{[\beta}\delta^\gamma_{\delta]}
-\delta^{[\alpha}_{[\beta}\delta^{\gamma]}_{\delta]}
\right].
\end{align}
Substituting Eq.(\ref{23a}) in (\ref{55a}), the Riemann
tensor is split as
\begin{equation}\nonumber
R^{\alpha\gamma}_{~~\beta\delta}=R^{\alpha\gamma}_{(I)\beta\delta}+R^{\alpha\gamma}_{(II)\beta\delta}+R^{\alpha\gamma}_{(III)\beta\delta},
\end{equation}
where, we have
\begin{align}\nonumber
R^{\alpha\gamma}_{(I)\beta\delta}&=2\kappa\left(\tilde{\rho}+\frac{s^2}{8\pi
r^4}\right)u^{[\alpha}_{[\beta}u^{\gamma]}_{\delta]}-2\kappa
\left(\check{P}+\frac{s^2}{24\pi
r^4}\right)h^{[\alpha}_{[\beta}u^{\gamma]}_{\delta]}+\kappa
(\tilde{\rho}-3\check{P})\\\nonumber&\times\left[\frac{1}{3}\delta^\alpha_{[\beta}
\delta^\gamma_{\delta]}-\delta^{[\alpha}_{[\beta}\delta^{\gamma]}_{\delta]}\right],\\\nonumber
R^{\alpha\gamma}_{(II)\beta\delta}&=2\kappa\left(\Pi-\frac{s^2}{4\pi
r^4}\right)\left[s^{[\alpha}s_{[\beta}\delta^{\gamma}_{\delta}+\frac{1}{3}
h^{[\alpha}_{[\beta}\delta^{\gamma]}_{\delta]}\right]+2\kappa\left[\tilde{q}s^{[\alpha}
u_{[\beta}\delta^{\gamma]}_{\delta]}+\tilde{q}u^{[\alpha}s_{[\beta}\delta^{\gamma]}_{\delta]}\right],\\\nonumber
R^{\alpha\gamma}_{(III)\beta\delta}&=4u^{[\alpha}u_{[\beta}E^{\gamma]}_{\delta]}-\epsilon^{\alpha\gamma}_{~~\mu}
\epsilon_{\beta\delta\nu}E^{\mu\nu},
\end{align}
fulfilling the following relations
\begin{equation}\nonumber
\epsilon_{\alpha\gamma\beta}=u^\mu \eta_{\mu\alpha\gamma\beta},\quad
\epsilon_{\alpha\gamma\beta}u^\beta=0.
\end{equation}
The vanishing occurs as a consequence of zero magnetic part of the
Weyl tensor. Also, we have used \cite{8}
\begin{eqnarray}\nonumber
\epsilon^{\mu\gamma\nu}\epsilon_{\nu\alpha\beta}=u^\sigma u^\rho
\eta^{\mu\gamma\nu}_{\rho}\eta_{\sigma\nu\alpha\beta},\quad
\epsilon^{\mu\gamma\nu}\epsilon_{\nu\alpha\beta}=\delta^\gamma_\alpha
h^\mu_\beta-\delta^\mu_\alpha
h^\gamma_\beta+u_\alpha\left(u^\mu\delta^\gamma_\beta-\delta^\mu_\beta
u^\gamma\right).
\end{eqnarray}
Contracting $\mu$ with $\alpha$, we get
\begin{equation}\nonumber
\epsilon^{\mu\gamma\nu}\epsilon_{\nu\mu\beta}=-2h^\gamma_\beta.
\end{equation}
The explicit expressions for the tensors
$X_{\alpha\beta},Y_{\alpha\beta}$ and $Z_{\alpha\beta}$ are
manipulated with field equations as
\begin{eqnarray}\label{60a}
&&X_{\alpha\beta}=\frac{\kappa}{3}\left[\tilde{\rho}+\frac{s^2
\kappa}{8\pi
r^4}\right]h_{\alpha\beta}+\left[\frac{\kappa\Pi}{2}-\frac{\kappa
s^2}{8\pi
r^4}\right]\times\left(s_{\alpha}s_{\beta}+\frac{1}{3}h_{\alpha\beta}\right)-E_{\alpha\beta},\\\label{61a}
&&Y_{\alpha\beta}=\frac{\kappa}{6}\left[\tilde{\rho}
+3\check{P}+\frac{s^2}{4\pi
r^4}\right]h_{\alpha\beta}+\left[\frac{\kappa\Pi}{2}-\frac{\kappa
s^2}{8\pi r^4}\right]\times\left(s_{\alpha}s_{\beta}+\frac{1}{3}
h_{\alpha\beta}\right)+E_{\alpha\beta},\\\label{62a}
&&Z_{\alpha\beta}=\frac{\kappa}{2}\tilde{q} s^\mu
\epsilon_{\alpha\mu\beta}.
\end{eqnarray}
Bel superenergy denoted by $\bar{W}$ and super-Poynting vector
denoted by $\bar{P}_\alpha$ as defined earlier in \cite{8} are given
as
\begin{align}\nonumber
\bar{W}=\frac{1}{2}\left(X_{\alpha\beta}X^{\alpha\beta}
+Y_{\alpha\beta}Y^{\alpha\beta}\right)+Z_{\alpha\beta}Z^{\alpha\beta},\quad
\bar{P}_\alpha=\epsilon_{\alpha\beta\gamma}\left(Y^\gamma_\delta
Z^{\beta\delta}-X^\gamma_\delta Z^{\delta\beta}\right).
\end{align}
Substituting the values of the three tensors discussed above, we acquire
\begin{align}\nonumber
\bar{W}&=\frac{5\kappa^2\tilde{\rho}^2}{24}+\frac{\kappa^2\tilde{\rho}\check{P}}{4}+\frac{3\kappa^2
\check{P}}{8}+\frac{\kappa^2\Pi^2}{6}+\frac{2E^2}{3}+\frac{\kappa^2\tilde{\rho}s^2}{16\pi
r^4}-\frac{\kappa^2 s^2 \Pi}{12\pi r^4}\\\nonumber
&+\frac{\kappa^2 s^2 \check{P}}{16\pi r^4}+\frac{\kappa^2
s^4}{64\pi^2r^8}+\frac{\kappa^2 \tilde{q}^2}{2},
\\\nonumber \bar{P}_\alpha&=\left(\frac{\kappa^2 \tilde{q} \tilde{\rho}}{2}+\frac{\kappa^2\tilde{q}\tilde{P}_r}{2}\right)s_\alpha .
\end{align}
The last equation exhibits that if there is no heat flux, the
super-Poynting vector becomes zero.
Bel Robinson scalar $(W)$ can be defined by the relation
$W=E^{\alpha\beta}_{\alpha\beta}$. Since the magnetic part of the
Weyl tensor vanishes in our case, we attain $W=\frac{2E^2}{3}$. Consequently, we can have
\begin{align}\nonumber
\bar{W}-W&=\frac{5\kappa^2\tilde{\rho}^2}{24}+\frac{\kappa^2\tilde{\rho}\check{P}}{4}+\frac{3\kappa^2
\check{P}}{8}+\frac{\kappa^2\Pi^2}{6}+\frac{\kappa^2\tilde{\rho}s^2}{16\pi
r^4}-\frac{\kappa^2 s^2 \Pi}{12\pi r^4}\\\nonumber &+\frac{\kappa^2
s^2 \check{P}}{16\pi r^4}+\frac{\kappa^2
s^4}{64\pi^2r^8}+\frac{\kappa^2 \tilde{q}^2}{2}.
\end{align}

\subsection{Five Relevant Scalars}

Now, we shall formulate five scalars previously calculated in
\cite{7} which will further be utilized in
Eqs.(\ref{47a})-(\ref{54a}). We noticed that $X_{\alpha\beta}$ and
$Y_{\alpha\beta}$ can be split into two parts, i.e., trace and
tracefree scalar parts. Working out such parts for
$X_{\alpha\beta}$, we attain
\begin{equation}\nonumber
X_{\alpha\beta}=\frac{1}{3}Tr X h_{\alpha\beta}+X_{<\alpha\beta>},
\end{equation}
where $Tr X=X^\alpha_\alpha$. Also,
\begin{equation}\nonumber
X_{<\alpha\beta>}=h^\mu_\alpha h^\nu_\beta
\left(X_{\mu\nu}-\frac{1}{3} Tr X h_{\mu\nu}\right),
\end{equation}
with
\begin{equation}\label{65a}
Tr X=X_{T}=k\tilde{\rho}+\frac{\kappa s^2}{8\pi r^4}.
\end{equation}
Consequently, we can have
\begin{equation}\nonumber
X_{\alpha\beta}=X_{TF}\left(s_\alpha
s_\beta+\frac{h_{\alpha\beta}}{3}\right),
\end{equation}
where
\begin{equation}\label{67a}
X_{TF}\equiv\left(\frac{\kappa\Pi}{2}-\frac{s^2\kappa}{8\pi r^4}-E
\right).
\end{equation}
Following the same steps, we acquire trace and tracefree parts for
$Y_{\alpha\beta}$ as
\begin{equation}\label{68a}
Tr Y\equiv
Y_T=\frac{\kappa}{2}\left(\tilde{\rho}+3\tilde{P}_r-2\Pi+\frac{s^2}{4\pi
r^4}\right).
\end{equation}
We procure
\begin{equation}\nonumber
Y_{\alpha\beta}=Y_{TF}\left(s_\alpha
s_\beta+\frac{h_{\alpha\beta}}{3}\right),
\end{equation}
here
\begin{equation}\label{70a}
Y_{TF}\equiv\left(\frac{\kappa\Pi}{2}-\frac{s^2\kappa}{8\pi r^4}+E
\right).
\end{equation}
Making use of the explicit expression for $Z_{\alpha\beta}$, we
obtain another scalar function as
\begin{equation}\label{71a}
Z=\sqrt{Z_{\alpha\beta}Z^{\alpha\beta}}=\frac{\kappa
\tilde{q}}{\sqrt{2}}.
\end{equation}
We observe that in the presence of electromagnetic field, $X_{TF}$
and $Y_{TF}$ describe local pressure anisotropy along with an
additional quantity representing the inclusion of charge as given
below
\begin{equation}\label{72a}
\kappa\Pi-\frac{\kappa s^2}{4\pi r^4}=X_{TF}+Y_{TF}.
\end{equation}
Now using these five scalar functions
$(X_{T},X_{TF},Y_{T},Y_{TF},Z)$, we can rewrite Eqs.
(\ref{47a})-(\ref{54a}) as
\begin{align}\nonumber
&\frac{\kappa\tilde{\rho}^*}{2}+\frac{1}{3}\left[X_{T}+X_{TF}+Y_{T}+Y_{TF}\right]\theta=
\frac{1}{3}\left(\theta+\frac{\sigma}{2}\right)\left(X_{TF}+Y_{TF}+\frac{\kappa
s^2}{4\pi r^4}\right)-\frac{\sqrt{2}Z^\dag}{2}\\\label{73a}
&\quad\quad\quad\quad\quad\quad\quad\quad\quad\quad\quad\quad\quad\quad\quad-
\sqrt{2}\left[Za+\frac{Zs^1}{r}\right],\\\nonumber &\frac{\kappa
\tilde{P}_r^\dag}{2}+\frac{a}{3}\left[X_{T}+X_{TF}+Y_{T}+Y_{TF}\right]+
\frac{s^1}{r}\left(X_{TF}+Y_{TF}+\frac{\kappa s^2}{8\pi
r^4}\right)=\frac{\sqrt{2Z}}{3}(\sigma-4\theta)\\\label{74a}
&\quad\quad\quad\quad\quad\quad\quad\quad\quad\quad\quad\quad\quad\quad\:\:\:-\frac{\sqrt{2Z^*}}{2}+\frac{ss'e^{-\lambda}}{4\pi
r^4},\\\label{75a}
&\theta^*+\frac{\theta^2}{3}+\frac{\sigma^2}{6}-a^\dagger-a^2-\frac{2as^1}{r}=-Y_T,\\\label{76a}
&\left(\frac{\sigma}{2}+\theta\right)^\dagger=-\frac{3\sigma
s^1}{2r}+\frac{3\sqrt{2}Z}{2},\\\label{77a}
&a^\dagger+a^2+\frac{\sigma^*}{2}+\frac{\theta\sigma}{3}-\frac{as^1}{r}-\frac{\sigma^2}{12}=-Y_{TF},\\\label{78a}
&\frac{1}{3}\left[(Y_{T}+Y_{TF})-2X_{TF}+X_{T}\right]\left(\frac{\sigma}{2}+\theta\right)+\left(\frac{X_{T}}{2}
-X_{TF}\right)^*=-\frac{3s^1\sqrt{2}Z}{2r},\\\label{79a}
&\left(\frac{\kappa\tilde{\rho}}{2}-X_{TF}\right)^\dagger=\frac{3s^1
X_{TF}}{r}+\frac{\sqrt{2}Z}{2}\left(\frac{\sigma}{2}+\theta\right),\\\label{80a}
&\frac{3m}{r^3}=\frac{X_{T}}{2}-X_{TF}.
\end{align}
The Bel superenergy and super-Poynting vector take the following
form
\begin{align}\nonumber
&\bar{W}=\frac{1}{6}(X_{T}^2+Y_{T}^2)+\frac{1}{3}(X_{TF}^2+Y_{TF}^2)+Z^2,\\\nonumber
&\bar{P}_\alpha=\frac{\sqrt{2}Z}{3}(X_{T}+X_{TF}+Y_{T}+Y_{TF})s_\alpha.
\end{align}

\subsection{On the Physical Meaning of Structure Scalars }

Following the work of Herrera et al. \cite{8}, this section is
focused on the physical interpretation of five scalar functions
procured in the previous section. Clearly, the trace part $X_{T}$ of
$X_{\alpha\beta}$ deals with the homogeneous energy density of the
system. The scalar $Z$ administers the heat dissipative flux. The
tracefree parts $X_{TF}$ and $Y_{TF}$ administers the local pressure
anisotropy in the presence of charge. To comprehend the physical
interpretation of scalar $Y_{T}$ and $Y_{TF}$, we make use of
Eq.(\ref{45a}) and obtain the following result
\begin{align}\nonumber
m_T&=(m_T)_\Sigma
\left(\frac{r}{r_\Sigma}\right)^3-r^3\int_r^{r_\Sigma}\frac{e^{(\lambda-\nu)/2}}{2r}\ddot{\lambda}dr
+r^3\int_r^{r_\Sigma}\frac{e^{(\lambda+\nu)/2}}{r}\\\nonumber&\times\left[Y_{TF}+\frac{s^2
\kappa}{r^4}\right]dr
-r^3\int_r^{r_\Sigma}\frac{1}{r^3}\left(\frac{\kappa
e^{(\lambda+\nu)/2}s^2}{8\pi
r}\right)'dr+r^3\int_r^{r_\Sigma}\frac{3\kappa s^*}{8\pi r^4}dr.
\end{align}
Considering our proposed system to be in an equilibrium or
quasi-equilibrium state, we acquire the following expression for the
Tolman mass
\begin{equation}\nonumber
m_T=\frac{\kappa}{2}\int^{r_\Sigma}_0 r^2
e^{(\nu+\lambda)/2}(T^0_0-T^1_1-2T^2_2)dr,
\end{equation}
which can alternatively be written as
\begin{equation}\nonumber
m_T=\int^r_0 r^2 e^{(\nu+\lambda)/2}Y_Tdr.
\end{equation}
Thus, $Y_T$ can be used to describe the Tolman mass for a spherical
object.

\section{Static Spheres With Anisotropic Pressure}

This section takes into account only the static spherically
symmetric systems defined previously in \cite{8}. Considering the
line element defined in Eq.(\ref{1a}), we express three different
substitutes to it, each one exhibiting different static spheres in
terms of above mentioned structure scalars.

\subsection{First Alternative}

By making use of Eqs.(\ref{30a}) and (\ref{80a}), we obtain
\begin{equation}\nonumber
e^{-\lambda}=1-\frac{2m}{r}+\frac{s^2\kappa}{8\pi r^2}.
\end{equation}
For a static sphere, using Eq.(\ref{75a}) and (\ref{77a}), we may
define the gravitational acceleration as
\begin{equation}\label{87a}
a=\frac{r}{3s^1}(Y_{TF}+Y_T).
\end{equation}
Alternatively, using Eqs.(\ref{21a}) and (\ref{22a}), we attain
\begin{equation}\label{88a}
a=\frac{e^{-\lambda/2}\nu'}{2},\quad s^1=e^{-\lambda/2}.
\end{equation}
Substituting Eq.(\ref{88a}) into (\ref{87a}) and integrating
w.r.t. radial coordinate, we get
\begin{equation}\nonumber
e^\nu=Ce^{\int\frac{2r}{3}(Y_{TF}+Y_T)\left[1-\frac{2r^2}{3}\left(\frac{X_T}{2}-X_{TF}+\frac{3\kappa
s^2}{16\pi r^4}\right)\right]^{-1}(Y_{TF}+Y_T)dr},
\end{equation}
with $C$ being the integration constant which one can easily determine from Eq.
(\ref{25a}). Consequently, in the static case, the line element
becomes
\begin{align}\nonumber
ds^2&=Ce^{\int\frac{2r}{3}(Y_{TF}+Y_T)\left[1-\frac{2r^2}{3}\left(\frac{X_T}{2}-X_{TF}+\frac{3\kappa
s^2}{16\pi r^4}\right)\right]^{-1}(Y_{TF}+Y_T)dr}dt^2\\\nonumber
&-\left[1-\frac{2r^2}{3}\left(\frac{X_T}{2}-X_{TF}+\frac{3\kappa
s^2}{16\pi r^4}\right)\right]^{-1}dr^2-r^2 sin^2 \theta d\phi^2.
\end{align}
It is worth noting that all the static anisotropic fluid spheres can
be fully expressed using the scalar functions $\frac{X_T}{2}-X_{TF}$
and $Y_{TF}+Y_T$.

\subsection{Second Alternative}

By making use of Eqs.(\ref{32a}), (\ref{65a})
and (\ref{80a}), we acquire
\begin{equation}\nonumber
m(r)=\frac{r^3}{3}\left[\frac{m'}{r^2}-\frac{1}{r^2}\left(\frac{\kappa
s^2}{16\pi r}\right)'-X_{TF}\right]+\frac{\kappa s^2}{16\pi r}.
\end{equation}
Performing integration w.r.t. $r$, we obtain
\begin{equation}\nonumber
m(r)=r^3\left(\int\frac{X_{TF}}{r}dr+C_1\right)+\frac{\kappa
s^2}{16\pi r}.
\end{equation}
Further, by making use of Eq.(\ref{30a}), we can write
\begin{equation}\nonumber\emph{}
e^{-\lambda}=1-2r^2\left(\int\frac{X_{TF}}{r}dr+C_1\right),
\end{equation}
where the constant of integration $C_1$ can be easily determined
from Eq.(\ref{25a}). Considering the static regime, the field
equation (\ref{9a}) yields
\begin{equation}\nonumber
\frac{\kappa
P_r}{2}=\frac{1}{2}\left(\frac{e^{-\lambda}-1}{r^2}\right)+\frac{\nu'e^{-\lambda}}{2r}+\frac{\kappa
s^2}{16\pi r^4},
\end{equation}
which on utilizing Eqs.(\ref{30a}), (\ref{68a}), (\ref{70a}),
(\ref{87a}) and (\ref{88a}), may be re-written as
\begin{equation}\nonumber
\frac{\kappa
P_r}{2}+\frac{m}{r^3}=\frac{\nu'e^{-\lambda}}{2r}+\frac{\kappa
s^2}{8\pi r^4}=Y_h,
\end{equation}
where
\begin{equation}\nonumber
Y_h=\frac{1}{3}\left(Y_T+Y_{TF}\right)+\frac{\kappa s^2}{8\pi r^4}.
\end{equation}
Upon integration, we found
\begin{equation}\nonumber
e^\nu=C_2e^{\int2rY_h\left[1-2r^2\int\frac{X_{TF}}{r}dr+C_1\right]^{-1}dr}.
\end{equation}
Here, again, $C_2$ symbolizes the integration constant which can be easily
attainable from Eq.(\ref{25a}). The line element under these
considerations takes the following form
\begin{align}\nonumber\emph{}
ds^2=C_2e^{\int2rY_h\left[1-2r^2\int\frac{X_{TF}}{r}dr+C_1\right]^{-1}dr}dt^2
-\left[1-2r^2\int\frac{X_{TF}}{r}dr+C_1\right]^{-1}dr^2-r^2 sin^2 \theta d\phi^2.
\end{align}
From the last equation, we conclude that all the possible space-times
exhibiting static anisotropic spheres can be expressed in terms of
scalar functions $X_{TF}$ and $Y_h$.

\subsection{Third Alternative}

Here, we workout the line element which is expressible in terms of
only the trace-free parts $X_{TF}$ and $Y_{TF}$ of $X_{\alpha\beta}$
and $Y_{\alpha\beta}$ respectively. Since we know that in the static
case, we have
\begin{equation}\label{94a}
E=-\frac{e^{-\lambda}}{2}\left[\frac{\nu''}{2}+\left(\frac{\nu'}{2}\right)^2+
\frac{\nu'}{2}\left(-\frac{\lambda'}{2}-\frac{1}{r}\right)+\frac{\lambda'}{2r}+
\frac{1-e^\lambda}{r^2}\right].
\end{equation}
Now, we introduce two new variables $y$ and $u$ as follows
\begin{equation}\nonumber
y=e^{-\lambda}, \quad \frac{\nu'}{2}=\frac{u'}{u}.
\end{equation}
Eq. (\ref{94a}) then takes the following form
\begin{equation}\nonumber
y'+2y\left(\frac{u''-\frac{u'}{r}+\frac{u}{r^2}}{u'-\frac{u}{r}}\right)=\frac{2u(1-2r^2E)}
{r^2\left(u'-\frac{u}{r}\right)}.
\end{equation}
Performing integration w.r.t. $r$, we acquire
\begin{equation}\label{95a}
y=e^{-\int k(r)dr}\left(\int e^{-\int k(r)dr}f(r)dr+c_1\right),
\end{equation}
where, we have
\begin{equation}\nonumber
k(r)=2 \frac{d}{dr}\left[ln\left(u'-\frac{u}{r}\right)\right],
\quad f(r)=\frac{2u(1-2r^2E)} {r^2\left(u'-\frac{u}{r}\right)},
\end{equation}
with $c_1$ being the constant of integration easily obtainable from
junction conditions. In terms of original variables, we can have
\begin{equation}\nonumber
\frac{\nu'}{2}-\frac{1}{r}=\frac{e^{\lambda/2}}{r}\times\sqrt{(1-2Er^2)+c_1r^2e^{-\nu}+
r^2e^{-\nu}\int\frac{e^\nu}{r^2}(2r^2E)'dr}.
\end{equation}
Now, introducing a new variable $z$ as follows
\begin{equation}\nonumber
e^\nu=\frac{e^{2\int zdr}}{r^2},
\end{equation}
from which we attain the following result
\begin{equation}\label{97a}
z(r)=\frac{\nu'}{2}+\frac{1}{r}.
\end{equation}
Making use of Eqs.(\ref{95a}) and (\ref{97a}), the newly introduced
variable $z$ and the Weyl tensor $E$ are interlinked as
\begin{equation}\label{98a}
z(r)=\frac{2}{r}+\frac{e^{\lambda/2}}{r}\sqrt{(1-2Er^2)+c_1r^4e^{-\int
2z(r)dr}+ r^4e^{-\int 2z(r)dr}\int\frac{e^{-\int
2z(r)dr}}{r^4}(2r^2E)'dr}.
\end{equation}
Next, utilizing field equations, we acquire
\begin{align}\nonumber
\kappa\Pi-\frac{\kappa s^2}{4\pi
r^4}=e^{-\lambda}\left[-\frac{\nu''}{2}-\left(\frac{\nu'}{2}\right)^2+
\frac{\nu'}{2r}+\frac{1}{r^2}\right]+\frac{\lambda'e^{-\lambda}}{2}\left[\frac{\nu'}{2}+
\frac{1}{r}\right]-\frac{1}{r^2}.
\end{align}
Using the new variables $y$ and $z$, we attain
\begin{align}\nonumber
y'+y\left[\frac{2z'}{z}+2z-\frac{6}{r}+\frac{4}{r^2z}\right]=-\frac{2}{z}\left[\kappa\Pi-\frac{\kappa
s^2}{4\pi r^4}+\frac{1}{r^2}\right].
\end{align}
Performing integration w.r.t $r$, the value of $\lambda$ comes out
to be
\begin{align}\label{99a}
e^{\lambda(r)}=\frac{z^2
e^{\int\left(2z+\frac{4}{zr^2}\right)dr}}{r^6\left[-2\int\frac{z}{r^8}\left(\kappa\Pi
r^2-\frac{\kappa s^2}{4\pi
r^2}+1\right)e^{\int\left(2z+\frac{4}{zr^2}\right)dr}dr+C\right]},
\end{align}
where $C$ symbolizes integration constant. Using the scalars
$X_{TF}$ and $Y_{TF}$, Eqs.(\ref{98a}) and (\ref{99a}) can also be
written as
\begin{align}\label{100a}
z(r)=\frac{2}{r}+\frac{e^{\lambda/2}}{r}\sqrt{[1-r^2(Y_{TF}-X_{TF})]+
r^4e^{-\int2zdr}\left(c_1+\int[r^2(Y_{TF}-X_{TF})]'\frac{e^{\int2zdr}}{r^4}\right)},
\end{align}
and for $\lambda$, we can have
\begin{equation}\nonumber
e^{\lambda(r)}=\frac{z^2
e^{\int\left(2z+\frac{4}{zr^2}\right)dr}}{r^6\left[-2\int\frac{z}{r^8}
\left(1+r^2(Y_{TF}+X_{TF})\right)e^{\int\left(2z+\frac{4}{zr^2}\right)dr}dr+C\right]}.
\end{equation}
If we consider conformally flat fluids having anisotropic pressure,
then $Y_{TF}=X_{TF}$. Making use of this condition in Eq.
(\ref{100a}), the value of $z$ takes the following form
\begin{equation}\nonumber
z=\frac{2}{r}+\frac{e^{\lambda/2}}{r}
\tanh\left(\int\frac{e^{\lambda/2}}{r}dr\right).
\end{equation}

\section{Conclusion}

Keeping in view the presence of electromagnetic field, a detailed
study relating to self-gravitating dissipative spherically symmetric
fluid is presented seeking the help of structure scalars that appear
when we split the Riemann tensor orthogonally. We found five such
scalars i.e., $(X_{T},X_{TF},Y_{T},Y_{TF},Z)$ which reduces to two
in number if we consider static and dissipation-less dust fluid with
anisotropic stresses and only one for the case of static isotropic
fluid distribution. We observe that $Z$ and $X_T$ delineate the
dissipative flux and the energy density respectively. For
dissipation-less fluid, the scalar $X_{TF}$ administers the energy
density inhomogeneity. The scalars $Y_{T}$ and $Y_{TF}$ appear in
the definition of Tolman mass with $Y_{TF}$ delineating the
consequences of inhomogeneity of energy density and anisotropy of
pressure on the Tolman mass of the proposed system and $Y_{T}$
delineating the Tolman mass density. Considering the static case, we
manipulated the Einstein equations as three ordinary differential
equations with five unknowns. We have observed that the solutions of
the field equations can be completely characterized by these scalar
functions in the presence of electromagnetic field like the charge
free case. Some particular solutions have been illustrated to
comprehend this argument and to signify the physical relevance of
the obtained scalar functions.

We have found some solutions to Einstein-Maxwell equations in terms
of structure scalars obtained from the splitting of the Riemann
tensor representing static anisotropic spheres. It predicts about
the formation and structure of stellar configurations including the
fact that the electromagnetic field results in an increase in the
mass of the static sphere (as predicted by set of equations in
second alternative). From the astrophysical point of view, we have
evaluated few results that are known to predict astrophysical
activities taking place in the cosmos. One of the best examples is
Tolman mass. Since, we know that expression for Tolman mass of a
stellar system provides information about the total energy budget of
that system, thus, the occurrence of the structure scalar $Y_{TF}$
in our calculated expression manifests that the presence of charge
does affect the total energy of any stellar object. Moreover, the
structural equations obtained above provide information about the
formation, structure and gradual development of stellar objects.
Structure scalars $(X_T, Y_T, X_{TF}, Y_{TF},Z)$ appearing in these
equations show that certain stellar configurations may exist that
have electromagnetic fields around them. These results depict the
effects of electromagnetic field on the structure of compact
objects. So, we can say that the stellar equations we obtained using
this formalism may represent the mathematics behind any static
anisotropic astrophysical object out there in the cosmos.

\vspace{0.25cm}

\noindent{\bf Acknowledgments}

\vspace{0.25cm}

\noindent The work of M. Z. Bhatti (PI) has been supported by
National Research Project for Universities (NRPU), Higher Education Commission,
Pakistan under research project No. 8769/Punjab/ NRPU/R$\&$D/HEC/2017.
Also, the authors would like to thank the Higher Education Commission, Islamabad, Pakistan for its
financial support through the {\it Indigenous Ph.D. Fellowship For 5000 Scholars, Phase-II, Batch-V}.

\end{document}